\newcommand{\be}{\begin{equation}}
\newcommand{\ee}{\end{equation}}
\newcommand{\bea}{\begin{eqnarray}}
\newcommand{\eea}{\end{eqnarray}}
\newcommand{\beas}{\begin{eqnarray*}}
\newcommand{\eeas}{\end{eqnarray*}}

\documentclass[onecolumn,aps,showpacs,nofootinbib,epsfig,axodraw]{revtex4}
\usepackage{graphics}
\usepackage{epsfig}
\usepackage{amsmath,amssymb}
\usepackage{booktabs}
\usepackage{xcolor}
\usepackage{multirow}
\usepackage{diagbox}

\begin{document}

\hfill CIFFU-03-17

\title{Where have all the large Representations gone? }
\author{J. Lorenzo Diaz-Cruz$^{1,2}$}
\affiliation{ $^1$ Centro Internacional de Fisica Fundamental, BUAP \\
$^2$ Facultad de Ciencias Fisco-Matematicas, \\
Benemerita Universidad Autonoma de Puebla,\\ 
Puebla, Mexico. 
}

\begin{abstract}

Gauge theories describe the interactions of the fundamental building blocks of nature with great success. 
The Standard Model achieves a partial unification of the electromagnetic and weak interactions, and
it also acomodates the strong interactions.  
The known quarks and leptons appear in the fundamental representations
(or singlets) of the $SU(3)_c\times SU(2)_L \times U(1)_Y$ gauge symmetry.
However, larger representations (EW triplets, color sextes, etc.) could also occur in
principle. Bounds on such exotic states based on electroweak precision tests, unitarity, perturbativity and 
collider searches, indicate that they should be very heavy or may be non-existent.
But why only small representations occur in nature?
Several ideas that could give some light into this problem are discussed here, 
including the approach of Nielsen et al, as well as the possible compositeness of  quarks and leptons.  
Then, we discuss the problem  within the context of grand unified theories, where  
a principle of "minimal complexity" is proposed to restrict the size of large representations, 
when they are required to form  unified multiplets. 
 
\end{abstract}


\maketitle

\section{Introduction}

The success of the Standar Model (SM) relies on a plethora of experimental tests,
which include now the discovery of the Higgs boson at the LHC collider experiments
\cite{higgs-atlas:2012gk,higgs-cms:2012gu}. The Higgs boson  mass ($m_h=125$ GeV)  is  consistent 
with  Electro-Weak (EW) precision measurements, which were tested previously by the  detection of the top quark, 
also with a mass lying in range expected from those precision tests \cite{Olive:2016xmw}.   
 These  results seems to confirm that the fundamental building blocks of nature and their interactions
are well described within the framework of quantum field theory (QFT).

However, the merits of the SM are somehow obscured by its arbitrary number of 
free parameters (19, without considering neutrino masses and mixing), 
which seem to be large for a theory that us supposed  to be a fundamental one.
Besides the gauge couplings, we have the Higgs parameters which are associated with the 
spontaneous breaking of the electroweak symmetry \cite{Gunion:1989we}, while the Yukawa parameters
trigger the generation of fermion masses.  Although the gauge couplings could 
be unified at the GUT mass scale, with some degree of success, there remmains the 
$\theta$ parameter, which in itself poses a problem \cite{Diaz-Cruz:2016pmm}. 
The origin of these SM parameters  is not known, 
despite the theoretical efforts from the last 30 years,
such as supersymmetry, extra dimensions and composite models \cite{Ellis:2010wx,DiazCruz:2004ss}.
Other problems that suggest the need for some extension of the SM include:
the lack of explanations for dark matter, baryogenesis, inflation, 
which can not be explained within the minimal SM \cite{Ellis:2017zwq}.

However, the searches for direct signals of new physics have resulted into
bounds on the corresponding mass scale that are now entering into the multi-TeV range.
Although we need to wait for future LHC stages in order to 
get stronger bounds on the new physics scale, and to reach a more solid conclusion  
on what comes after the SM, the fact that the LHC has not found evidence of new physics
is perhaps telling us something important. May be we should start thinking 
at a deeper level and to scrutinize the SM structure, in order to find clues that may
allow us to understand those open problems of the SM.

One such regularities of the SM, but for which we do not have an explanation, 
is the appearance of matter in representation  of small size. Namely, after decades of experimental 
guide, ranging from beta-decay to  the discovery of top quark, we know that matter appears only in 
certain representations of the  gauge symmetries, 
namely the known quarks and leptons are either singlets or appear in the fundamental
representations (doublets  and/or triplets) of $SU(2)_L$ and $SU(3)_c$, respectively. 
Furthermore, the chiral fermions must have different properties,  in order to fit the maximal parity 
violation observed in the weak interactions.
Is this whole arrangement just something given? or is it telling us somethin deep? Is it an issue 
to ask where have all the large representations gone?

Furthermore, one can classify the fundamental constituents of the visible universe by tabulating
the spin ($S$) and weak isospin ($T$) that could occur in a generic QFT, as shown in table 1. It seems remarkable that 
the whole (visible) universe is built out of some very few entries of this table.  Although some of these states 
are realized in nature, there are other possibilities that have not been detected yet, despite having 
some strong theoretical motivations; such as the spin-2 ($T=0$) graviton, which appears in case the gravitational 
force were quantized as the other fundamental interactions of nature. Considerations based on supersymmetry
provide  motivations for the existence of its spin-3/2 superpartner (also with $T=0$),  the gravitino
\cite{Dudas:2017rpa}.


\begin{table}[t!] 
\begin{center}  
\begin{tabular}{|| c|| c| c| c| c| c||}  
\hline\hline
     \diagbox{\textbf{T}}{\textbf{S}} &  0   &    $\frac{1}{2}$    &  \,  1  \,    &  \,  $\frac{3}{2}$ \,  &  \, 2 \, \\
     \hline \hline 
     0          &    ?    &  $q_R,l_R$ &  Gluon   &  ??  &   ?? \\
                &         &    $\nu_R$(?)    &          &     &     \\
     \hline  
  $\frac{1}{2}$ &   Higgs   &  $Q_L,L_L$    &  ?   &   ??     &   ?? \\
                &          &       &        &        &   \\
\hline  
     1    &   ?      &    ?    &     $W, Z$ &   ??     &  ??  \\
            &         &        &          &     &     \\
\hline  
 \hline 
 \end{tabular}  
\end{center}  
\end{table}

In this paper, we shall assume that the appearance of small repreesentations within the SM 
needs some explanation, and shall discuss several aspects of the problem. First, in section 2 
we discuss the known theoretical and experimental bounds on the mass scale of exotic representations, 
including electroweak precision tests, unitarity, perturbativity and collider searches.
Then, in section 3 we discuss some ideas aimed to solve this problem, which we label as
{\it{effective selection rules}}, here we include the work of  H. Nielsen et al. \cite{Nielsen:2015lpa},  
as well as the use of spacetime discrete symmetries; we close the section with a discussion of the implications 
on the representation problem from a possible sub-structure of quarks and leptons.
Then, in section 4,  we discuss how to restrict the matter representatios following arguments
based on the unification paradigm. We postulate a  "Principle of Minimal Complexity"  
which could help to discriminate the allowed representations in nature, includig the Higgs multiplets,
whenever they are required to forma unified multiplets.
Finally, section 5 contains our conclusions and outlook.

\section{Constraints on Large Representations}

The fact  that only a restricted set of representations of the gauge symmetry is realized in nature,
suggest that some fundamental principle is missing in our current formulation of the SM; 
something that will tell us why nature only uses a few entries of the 
Table 1. The study of the properties and phenomenology of the missing large representations,
provide some limits on its mass scale. 
Here, we shall consider the constraints derived from the following arguments:

\begin{enumerate}

\item {\bf{Electroweak Precision tests (EWPT)}}  The indirect effects of heavy quanta
on ohysical observabes, is conveniently represented by the Peskin-Takeuchi parameters ($S,T,U$), 
which has been treated in  many works and applied to several cases. 
In particular, Ref. \cite{Zhang:2006de} contains  the evaluation of the effects from 
higher-dimensional fermion representations on the S, T,U parameters. It is found
that in order  to keep $T$  UV finite one needs to imposse a sum rule for the states  with 
isospin $j,j_3=l$ and corresponding mass $m^2_j$, i.e.:
\begin{equation}
\sum_l (j^2+j-3l^2)m^2_j=0,  
\end{equation}
This relation is satisfied automatically for a weak doublet ($T=1/2$), but for general isospin it imposses
constraints on the mass of the multiplt components, which look somehow ad hoc. In fact, this result could
be already used to look at the large representations with suspicius eye.

\item {\bf{Unitarity bounds}}. Invoking the unitarity bound has been a useful tool in order to analyze the high-energy 
behavior of a given QFT. Here, one requires that the zeroth partial wave amplitude ($a_0$) satisfies the 
tree-level partial wave unitarity, i.e. $|Re(a_0)| \leq 1/2$. The amplitudes are evaluated useing perturbative
methods from the corresponding lagragian.

The authors of Ref. \cite{Hally:2012pu}, find that in order to respect the unitarity argument
the weak isospin of a complex scalar multiplet must satisfy the bound:  
$T \leq \frac{7}{2}$.
They have also combined the EWPT with the unitary bounds for particular models that include an stable DM
candidate, to further constrain the allowed parameter space for the large
representations. In particular, from their figure 1, one reads that
small mass difference between the stable component of the mutiplet and the first charged one, are allowed;
for instance, taking $\Delta m = 5$ GeV and $M_{dm} \simeq 5$ TeV, implies that values of $j>7/2$ are 
already excluded. 

\item {\bf{Perturbativity}}. In Ref. \cite{Cirelli:2005uq}, the authors discuss bounds on the weak isospin,
obtained from analyzing the contribution of such large representations to the RG evolution of the $SU(2)_L$
gauge coupling. They find that a real multiplet must have $T\leq 3$ in order to avoid $g_2$ entering into the
non-perturbative domain, below the Planck scale. For a complex multiplet, they find that the
corresponding limit becomes $T\leq 5/2$. 

\item {\bf{Collider searches}}. The LHC has presented results from dedicated searches for 
exotic particles. In some cases, they present model-independent results for specific states 
(such as leptoquarks),  while in other cases  the searched particles are part of well motivated theories.
For instance the search for gluinos within the MSSM can also be considered as the search for a colored fermion
octet. Overall, the LHC bound for such states is of order $M>O(1)$ TeV.

\end{enumerate}

Thus, we conclude that large representations, beyond the SM ones, are disfavored by present data.
 Although those results do not constitute a proof to
forbidde its existence, it certainly motivates the need for an explanation of the absence of
such exotic representations.
If such states are not detected in the future LHC runs, it will become more pressing to look for an
explanation of this aspect of the SM.

\section{Effective selection rules and compositeness}

\subsection{What do we know? Effective selection rules.}

For some time, H. Nielsen and collaborators \cite{Nielsen:2015lpa}, have studied how to make sense on the
peculiarities of the SM structure, using Group Theory and some physical arguments.  We shall call their approach
as "effective selection rules", in the sense that some efffective explanation is seeked, which may be written using 
only the SM  degrees of freedom, which are assumed provisionaly to be fundamental ones, although 
they may actually be of some emergent nature, at a deeper level. 
 Nielsen and collaborators have emphasized that some aspects of the SM structure
, e.g. charge quantization among others, can be better understood by distinguishing between the Lie algebra
of the SM, i.e. $u(1)_Y\times su(2)\times su(3)$, and the Lie group, which happens to be $S(U(2)\times U(3))$,
thanks to the structure of the corrresponding covering group, and the relationship between
the hypercharge and the duality/triality of the fermion representations.
Furthermore, they also show that 
the SM gauge group is rather special, as it shows a minimal degree of "skeweness", 
namely that it has the smallest number of automorphisms \cite{Nielsen:1989na}. 

In order to quantify the problem of the absence of the large representations, Nielsen et al. 
have built an argument to show that quarks and leptons appear not only in the smallest representations
of the non-abelian $SU(2)$ and $SU(3)$ gauge groups, but that also the abelian charges are
of minimal size. Their arguments involves a combined  use of Han-Nambu charges and
anomaly cancellations. 
More recently, they have argued that not only the matter representations fall into the fundamental representations,
but also the gauge groups has associated some quantity that gets maximized for the SM 
gauge group \cite{Nielsen:2014gfa}. This  quantity is defined as a modification of 
the ratio of the quadratic Casimir invariant for the adjoint representation and that for the samallest
faithfull representation.  This quantity  gets maximized and singles out the $S(U(2)\times U(3)$ 
gauge group, which is singled out. 
Furthermore, this quantity also helps to single out the fundamental representation for the quarks, leptons and 
even the Higgs doublets.

Another line of reasoning has been presented in ref.  \cite{Bento:1991zu}, which uses discrete symmetries 
as a way to constrain  the size of the allowed represenntations. First, it is argued that only
some representations allow to use consistently the same definition for the
discrete symmetries. Then, by using invariance under change of basis restricts possible representations,
it is concluded that only fundamental representations of SU(N) are allowed.
In this regard, we should also mention that  Nielsen et al. have extended their work to
look for a related quantity that can be applied to the dimension of spacetime ($D$), and they
find that such quantity gets maximized for $D=4$.

We could also mention that a similar discussion holds for the allowed spin that can be described
consistently within QFT. In that case,some difficulties are associated with the propagation of higher 
spin fields, when they interact with an external c-number  field, which may include problems with
causality, the presence of unphysical states and violation of unitarity. These results indicate that
fundamental particles with spin $S \leq 2$ have difficulties to be treated in QFT \cite{Weinberg:1995mt}.
If somehow the isospin were related with the spin, it may be possible to argue 
that only the related small values of isospin for fundamental particles could be included in QFT. 
One such possibility could be the extra dimensional scenarios, 
where the interactions have a geometrical origin.

\subsection{Small representations from compositeness}

Based on the successive sub-structure layers that appear when going from molecules to atoms, and from there
to nuclei and hadrons, we could also argue that something similar could
happen at the quarks and leptons level. This would explain why we only observe small representations,
due to a dynamical reason. Namely, if the known quarks and leptons were 
composite, the dynamics could operate in such a way that it does not allow  the appearence
of larger representations. 

It could be usefull to review the similar situation  in nuclear and hadron physics, to illustrate 
our point and to gain some insight into the problem.
Namely, we can recall that only nuclei with atomic number $Z \lesssim 120$ are observed in nature. 
We know that this happens because  in this limit the nuclear attractive force between nucleons ($p\,\&\,n$) 
dominates over the electromagnetic repulsion. 
Namely, when 
\begin{equation}
 F_{em}\sim k\frac{Z^2 e^2}{r^2} \sim F_{nucl} \simeq \frac{e^{-mr}}{r}, 
\end{equation}
Then one finds that this implies: 
$Z\simeq 1/\alpha_{em} \simeq 137$. 

Similarly, it could happen that quarks and leptons are composite, but in such a way that their quantum numbers 
only allow doublets of $SU(2)_L$ and triplets of $SU(3)_c$, because there exist a repulsive force among
the constituents that makes impossible to form bound states with large values of isospin or color.
If this were the case, it is possible that some imprint would be left on the SM fermion properties 
by the composite dynamics, for instance on the anomalous magnetic coments.
In ref. \cite{Dai:2001vv},  several models were proposed to treat the anomalous
magnetic moment of a composite fermion, with an scale $\Lambda \simeq 1$ TeV.

\section{SM representations and Grand Unification.}

Another possible argument to explain the appearence in nature
of matter fermions in small representation, could arise within the unification paradigm.
Namely, by assuming that nature only accepts the known SM multiplets or at most the
addition of  minimal extensions, we could ask what are the allowed dimensions for such representations,
when they are assumed to form unified mmultiplets under some unified gauge group.

\subsection{GUTs and the principle of "Minimal Complexity" }

Let us discuss our point within the context of minimal $SU(5)$ GUT. Here one finds the 
remarkable property that the known quarks and leptons, being  color triplets or weak doublets (singlets), 
can fit into an (almost) small unified representation, e.g. $5$ and $10$ of $SU(5)$. Both of these
representations can also be accomodated into the  spinorial $16$ dimensional representation of  $SO(10)$ GUT,
 provided that a righ-handed singet neutrino is added to the SM matter fields.

Then, we could ask whether the addition of extra multiplets, charged under the SM gauged Lie Algebra, 
could still be unified into sthe gauge groupunder consideration, without adding much complications. It turns out 
that this is not an easy task, because as soon as one include some larger weak-color representations,
one needs to invoke  larger GUT multiplets.

As we do not like that, we shall look for an argument  to keep thedimension of the unified 
multiplets as small as possible. Thus, in order to achieve that, we present a principle designed 
to restrict the dimensions of the representations allowed in some given GUT, which we shall call 
"Principle of Minimal Complexity", which says the following: 

\bigskip

{\large{{\it{ 
{\textcolor{purple}{"When a multiplet of certain dimension is added tothe  SM, no representation of larger dimension 
should be allowed  in order to  have a complete multiplet under some GUT gauge group"}}
}}
}}

\bigskip

We can apply this principle to discuss which  representations of larger dimensions (beyond doublets of $SU(2)_L$
and triplets of $SU(3)_c$) can have a natural place into the different theories. For instance,  electroweak triplets 
have been considered in the literature for several purposes, 
such as generating neutrino masses, getting the right Baryon asymmetry of the Universe and 
also to study Dark Matter candidates. The question is how natural is the inclusion of such triplets in
different GUT's? What is the prize that should be payed in terms of extra mutiplets?

For the case of $SU(5)$ GUT group, we show in table 2 the branching rules for the representations
of dimension $N=5,10,15, 24, 35, 45$. We can see from this table that adding a weak triplet forces the 
addition of extra  multiplets. 
For instance, we find that the triplet can be unified into $15$-dim. repr., but then besides the weak triplet, 
one needs to add a color sextet $(6)$. 
Similarly, when we consider $SO(10)$ GUT, breaking into the Pati-Salam model  
$SU(2)_L\times SU(2)_R\times  SU(4)$, the SM multiplets fit well into the $16$-dim representation.
But then the branching rules dictate that the inclusion of extra triplets would require invoking 
the $45$ or $54$ dimensional representations, which come with extra sextets too.
Thus, if we just want to add extra fermions, the safest way to form unfied multiplets, within $SU(5)$ 
is to add fermions that belong to the $N=5,10$ dimensional representations, which would then form the $16$ 
under $SO(10)$.

On the other hand, adding extra Weak fermions can also be considered within trinified models 
$SU(3)_c\times SU(3)_L \times SU(3)_R$. The SM fermions can  appear within the fundamental 
representation $(3,3,1)$ \cite{DiazCruz:2010dc}, but with extra fermions. Weak triplets can be accomodated
within an octet of $SU(3)$,  which decomposes under $SU(2)\times U(1)$ as:  $8=3+1+2+2$. In this case 
the multiplet does not involve representations larger than what has been included.  
Thus, adding extra fermions seem to be more natural within this
trinified model.

\begin{table}[t!] 
\begin{center}  
\begin{tabular}{|| c|| c|| c||}  
\hline\hline
     \textbf{Representations of} $\mathbf{SU(5)}$ &     $\mathbf{SU(2)} $     &      $\mathbf{SU(3)}$  \\
     \hline\hline
                {\multirow{2}{*}{$\mathbf{5}$} }                & $\mathbf{2}$ & $\mathbf{1}$\\ 
                				  & $\mathbf{1}$ & $\mathbf{3}$\\
				  				  \hline
                         {\multirow{3}{*}{$\mathbf{10}$} } & $\mathbf{1}$ & $\mathbf{1}$\\
				&   $\mathbf{1}$ & $\mathbf{\bar{3}}$ \\
				  & $\mathbf{2}$ & $\mathbf{3}$ \\
				  \hline
				   {\multirow{3}{*}{$\mathbf{15}$} } & $\mathbf{3}$ & $\mathbf{1}$\\
				&   $\mathbf{2}$ & $\mathbf{3}$ \\
				  &$ \mathbf{1} $& $\mathbf{6}$ \\
				  	  \hline
				   {\multirow{5}{*}{$\mathbf{24}$} } & $\mathbf{1}$ & $\mathbf{1}$\\
				&   $\mathbf{3}$ & $\mathbf{1}$ \\
				  &$ \mathbf{2} $& $\mathbf{3}$ \\
				   &$ \mathbf{2} $& $\mathbf{\bar{3}}$ \\
				    &$ \mathbf{1} $& $\mathbf{8}$ \\
				      \hline
				   {\multirow{6}{*}{$\mathbf{35}$} } & $\mathbf{4}$ & $\mathbf{1}$\\
				&   $\mathbf{3}$ & $\mathbf{\bar{3}}$ \\
				  &$ \mathbf{2} $& $\mathbf{\bar{6}}$ \\
				   &$ \mathbf{1} $& $\mathbf{\overline{10}}$ \\
				    &$ \mathbf{2} $& $\mathbf{1}$ \\
				     \hline
	{\multirow{5}{*}{$\mathbf{45}$} } &   $\mathbf{1}$ & $\mathbf{3}$ \\
				  &$ \mathbf{3} $& $\mathbf{3}$ \\
				   &$ \mathbf{1} $& $\mathbf{\bar{3}}$ \\
				    &$ \mathbf{1} $& $\mathbf{6}$ \\
				     &$ \mathbf{2} $& $\mathbf{8}$ \\
\hline  \hline 
\end{tabular}  
\end{center}  
\end{table}

A more ambitius unification program would
include the horizontal (family) symmetry. In such case, it could be interesting to look at the
constraints that some gauge-family unification could imposse to
restrict the size of the unified representations. 
 In this regard, it is worth mentioning the work of  \cite{Wilczek:1981iz},
which looked at the  properties of orthogonal Lie groups; their aim was  to look  for some explanation
of the SM structure \cite{Zee-kyoto}.
Starting from a great gauge group $\cal{G}$ that contains a single representation $\cal{R}$ that
accomodates the three families, one looks for its breaking into some  GUT gauge group $G$:  ${\cal{G}} \to G$,  
in such a way that the representation also break as: ${\cal{R}} \to R+R+R+...$,
with {\textcolor{cyan}{$R$}} being one family. 
It turns out that the group $SO(2(n+m))$ enjoy the property that spinor representations break into $2^m$ 
representations of $SO(2n)$,
For $n=5, m=4$, the group $SO(18)$ breaks by giving 8 families of $SO(10)$ 
(V-A) + mirror fermions (V+A).
 It could be interesting to search weather there exists some group ${\cal{G}}$ 
that it only allows a unique representation, but so far we have not found a satisfactory answer.


\subsection{Higgs Representations}

So far we have discussed the absense of large representations in the matter (fermionic) sector.
But a similar discussion could arise in the Higgs sector. There are plenty of motivations for
enlarging the scalar sector beyond the SM Higgs doublet, such as generating small neutrino
masses, hierarchy problem, dark matter, BAU, which include extra scalar multiplets. The extensions of
the Higgs sector could include extra singlets, doublets and triplets. Models with $T=3/2$ for neutrino masses
have been considered in the literature, whereas models with $T\geq 2$ have been shown to provide stable candidates
for dark matter \cite{Cirelli:2005uq}.

But even in such case, we could again ask what is the limit on the dimension of the 
allowed Higgs representations. When the neutral component of the Higgs multiplet of weak isospin $T$
and hypercharge $Y$, aquires a vev, it contributes to the $\rho$ parameter 
( $\rho = m^2_W /m^2_Z \cos^2 \theta_W$). In order to keep $\rho \simeq 1$, as it is observed in nature, 
one needs to have: $(2T+1)^2 = 3Y^2-1$ \cite{Gunion:1989we}. This relation is always satisfied for Higgs doublets, 
but one needs to imposse some tunning of the vevs whenever we consider a more exotic Higgs sector.

An additional problem arise when the Higgs sector is embedded into a SUSY theory, namely 
the Higgs scalars are acompanied by its fermionic partners, 
the Higgsinos, which in general provide a non-vanishing contribution to the abelian anomaly.
To have anomaly-free models usually a vector-like theory is choosen, i.e. including fermions of both
chirality which couldbe made very heavy. However, as shown in ref. \cite{Aranda:2009wh}, it is also possible 
to have theories of chiral type, which also satisfy the anomaly cancellation conditions. 

For instance, within $SU(5)$ GUT, we know that the representatios ${\bf 5}$ 
and  $\overline{\mathbf{10}} $ have opposite contributions to the gauge abelian anomaly, 
so together they form an anomaly free representation. As we said before,
sets of representations of $SU(5)$ that complete some representation of $SO(10)$ are anomaly free. 

But what about large representations, such as the $45$?. This case appears in some solutions to the 
problem of quark-lepton mass relations. 
In this case we could need extra representations in order to cancel anomalies.
The simplest choice is to consider a vector like SUSY Higgs sector, which includes the $45$ and $\overline{\mathbf{45}}$.
But this is not the only  choice; as shown in ref. \cite{Aranda:2009wh}, we could also consider
satisfying the anomaly cancellation condition using a combination of lower-dimensinal representations.
For instance, we can have a combination of multiplets that satisfy:
\begin{equation}
 A(45)+c_5 A(\bar{5}) + c_{10} A(\bar{10}) + c_{15} A(\bar{15})= 0
\end{equation}
where the anomaly coefficients are: $A(45)=6$, $A(\bar{5}) =-1$, $ A(\bar{10}) = -1$, 
$A(\bar{15}) = -9$. For instance a solution to the above equations is given by:
$c_5=c_{10}=3$ and $c_{15}=0$, plus plenty of other solutions.

But again, the introduction of larger dimensional multiplets usually involve the addition of
even large representations in vector-like theories, or a combination of several mutiplets of
lower dimension. Besides the aesthetic problems, the inclusion of such multiplets make the 
unified gauge coupling constant to run into the non-perturbative domain, before the Planck scale.

\section{Conclusions}

We have studied the problem of the SM representations, namely the fact that the 
quarks and leptons appear only in the fundamental (or singlet) representations 
of the $SU(3)\times SU(2) \times U(1)$ gauge symmetry. Larger representations 
(EW triplets, color sextes, etc.), which in principle could also occur in nature, 
shine for its absense. Namely, considerations based on EWPT, unitarity, perturbativity  
and collider bounds, provide bounds on the corresponding mass scale, which 
are entering into the multi-TeV domain. We have reviewed some approaches to 
understand the absence  of such representations,  including: {\it{effective selection rules}},
use of discrete and spacetime symmetries. An explanation for the problem of large representations, 
based on a possible substructure layer  for the known quarks and leptons, was also briefly 
considered here.

We have also discussed this problem  within the grand unification paradigm, here we  
proposed  a principle of "minimal complexity", which sates that when one adds some 
representations of certain size, one should not permit the addition of even  larger ones, 
in order to form  unified multiplets. As an application of this principle, we argued that, for instance,  
adding EW triplets is not economical  within the $SU(5)$ or $SO(10)$ models, 
but they could arise more naturally within a trinified model $SU(3)_c\times SU(3)_L \times SU(3)_R$.
Here one may question the validity of this approach when one does not include the flavor
problem, but then the problem gets really complicated.

May be some radical new approch would be needed in order to explain
 the apparent absence of large representations. Here I like to speculate that may be 
it is time to involve number theory into particle physics. For instance, it is 
amusing to identify the number of generators in the SM Lie algebra,
$SU(3)_C\times SU(2)_L \times U(1)_Y   $, i.e. the integers $3,2,1$, 
which are the factors of the first perfect number (6). In fact, the prime nature of these numbers,
appears in the discussion of  Nielsen et al. \cite{Nielsen:2014gfa}, in their attempt to  go beyond 
the usual considerations of the Lie Algebra of the SM, and actually identify the true gauge group 
of the SM, as a subgroup of the covering group, which is discussed in detail in Ref. \cite{ORaifert}.

\bigskip

{\bf Acknowledgements. Work supported by CONACYT, SNI (Mexico) and special grants from
VIEP-BUAP. Many thanks to A. Aranda, B. Larios and Ulises Saldana, for motivating discussions
on this topic. I also want to thank K. Tsumura for the opportunity to present these results as a talk at
Kyoto University.}

\end{document}